\documentclass[preprint,amsmath, amssymb, amsfonts, nofootinbib]{revtex4}
\usepackage{amsthm}
\usepackage[title]{appendix}
\newtheorem{thm}{Theorem}[section]
\newtheorem{lemma}[thm]{Lemma}
\newtheorem{prop}[thm]{Proposition}
\newtheorem{cor}[thm]{Corollary}
\newtheorem*{Uri}{Urysohn Lemma}
\theoremstyle{remark}
\newtheorem*{note}{Note}
\theoremstyle{definition}
\newtheorem{defn}[thm]{Definition}
\newtheorem{exam}[thm]{Example}
\newcommand{\eq}{\begin{equation}}
\newcommand{\eeq}{\end{equation}}
\newcommand{\natr}{\mathbb{N}}
\newcommand{\integ}{\mathbb{Z}}
\newcommand{\rat}{\mathbb{Q}}
\newcommand{\real}{\mathbb{R}}
\newcommand{\T}{\mathcal{T}}
\makeatletter
\DeclareRobustCommand\bfseries{%
  \not@math@alphabet\bfseries\mathbf
  \fontseries\bfdefault\selectfont\boldmath}%
\makeatother

\begin{document}
\title{Some results on the Zeeman topology}
%\date{}
\author{Giacomo Dossena}
\email{dossena@mat.unimi.it}
\affiliation{Dipartimento di Matematica, Universit\`a degli Studi di Milano, via Cesare Saldini 50, I-20133 Milano, Italy.}

\begin{abstract}
In a 1967 paper \cite{Zeeman}, Zeeman proposed a new topology for Min\-kow\-ski spacetime, physically motivated but much more complicated than the standard one. Here a detailed study is given of some properties of the Zeeman topology which had not been considered at the time. The general setting refers to Minkowski spacetime of any dimension $k+1$. In the special case $k=1$, a full characterization is obtained for the compact subsets of spacetime; moreover, the first homotopy group is shown to be nontrivial.
\end{abstract}

\maketitle
%\tableofcontents

\section{Introduction}
The Zeeman topology is a topology on Minkowski spacetime $M$, different from the standard one, introduced by Zeeman \cite{Zeeman} in 1967. The motivations for Zeeman's construction lie in the following facts: the standard topology does not account for the causal structure of spacetime, \emph{i.e.}, it does not distinguish among timelike, spacelike and lightlike directions; moreover, the homeomorphism group of $M$ endowed with the standard topology contains many elements with no straightforward physical meaning. Instead the Zeeman topology, defined to be the finest topology inducing the standard one on each straight timelike line and each spacelike hyperplane, enjoys the following properties: the homeomorphism group is generated by the Lorentz group, translations and dilatations; the light, time and space cones at a point are orbits under the action of the homeomorphism group.

The seminal work of Zeeman has inspired several generalizations, also applying to curved spacetimes, always based on the idea to use the causal structure to induce a topology. We will mention, in particular, the following contributions: G\"{o}bel \cite{Goebel1} \cite{Goebel2} defined and investigated, among other things, the general relativistic analogue of the Zeeman construction; Hawking, King and McCarthy \cite{HKM} proposed a topology, defined in terms of continuous timelike curves; Fullwood \cite{Fullwood} defined and studied an elegant topology, similar to the one of Alexandrov; Kim \cite{Kim} extended the construction of \cite{HKM} to the Budic-Sachs causal completion of a globally hyperbolic Lorentzian manifold.

In this paper we stick to the original work of Zeeman on flat spacetime. Our analysis will clarify some more or less known facts; we will also present some results which are new, to the best of our knowledge.

This paper begins with a presentation of the Zeeman topology, in the language of affine spaces (Sections \ref{sec:prelim} and \ref{sec:Zeeman_defined}). Sections \ref{sec:Zeno} and \ref{sec:1+1subsets} contain concepts and tools to be used in the sequel. In the other Sections, among other things, the Zeeman topology is proved to be separable, non-first-countable (Section \ref{sec:count}) and non-normal (Section \ref{sec:normality}). Other features of this topology are analyzed when $M$ has dimension 1+1; in particular, we show (Section \ref{sec:compact}) that a subset of $M$ is compact in the Zeeman topology if and only if it is a union of finitely many compact subsets with respect to the standard topology, each lying on a timelike or spacelike straight line. Still for the 1+1 dimensional case, we show (Section \ref{sec:homotopy}) that the first homotopy group of $M$ is not trivial, containing uncountably many subgroups isomorphic to $\integ$.

To our knowledge, the separability of the Zeeman topology has not been discussed in the literature so far; non-first-countability and non-normality were stated in \cite{Zeeman} without fully explicit proofs. The results on compact sets and on the first homotopy group in 1+1 dimension are presumably new.

\section{Preliminaries on affine spaces and Min\-kow\-ski spacetime}\label{sec:prelim}

\subsection{Motivations and terminology}
A precise description of the Zeeman and of other topologies on Minkowski spacetime $M$ must take into account its affine nature, \emph{i.e.}, distinguish between \emph{points} of $M$ and \emph{vectors} of the associated vector space $M_0$.

We take \cite{Kostrikin} as our standard reference on affine spaces. In general, if $F$ is an affine space, we write $F_0$ for the associated vector space; the sum of points and vectors is a map $+\colon F\times F_0\to F$, $(p,v)\mapsto p+v$. We call $F$ an affine space \emph{over} $F_0$. Of course, from any vector space $F_0$ we can make an affine space setting $F:=F_0$, and defining $+$ as the usual vector sum of $F_0$. Needless to say, $F_0$ is an affine space of a peculiar kind, since points and vectors coincide, and there is a distinguished point (the zero element).

Let $F$ be any affine space. A subset $S\subset F$ is an affine subspace if $S=p+S_0$ for some $p\in F$ and some vector subspace $S_0\subset F_0$. In an obvious way, $S$ becomes an affine space over the vector space $S_0$.

\subsection{Topologies on affine spaces}
Let $F$ be an affine space; the associated vector space $F_0$ could even be infinite dimensional. Let $\T$ be a topology on $F$; it is natural to ask $\T$ to somehow respect the affine structure of $F$. A possible way to formalize this requirement is the following:

\begin{defn}
$\T$ is \emph{compatible with translations} if, for each $v\in F_0$, the translation $T_v\colon F\to F$, $p\mapsto p+v$ is continuous.
\end{defn}

\begin{note}
Since $T_v^{-1}=T_{-v}$, equivalently $\T$ is compatible with translations if, for each $v\in F_0$, the translation $T_v$ is a homeomorphism.
\end{note}

Any $\T$ as in the previous definition comes in fact from a topology on $F_0$, in the following sense:

\begin{defn}
Given a topology $\T_0$ on $F_0$, the \emph{translated topology} on $F$ is:
\eq
\T:=\{\Omega\subset F \mid \forall p \in \Omega \;\exists O_p\in\T_0 \text{ such that } 0\in O_p \text{ and } p+O_p\subset \Omega\}\quad.
\eeq
\end{defn}

\begin{prop}\label{prop:top_affine}
For any topology $\T$ on $F$, the following facts are equivalent:
\begin{enumerate}
\item $\T$ is compatible with translations.
\item $\T$ is the translated topology of a topology $\T_0$ on $F_0$.
\end{enumerate}
Moreover, if \textit{1} holds, among the topologies $\T_0$ as in \textit{2} there is at most one compatible with translations of $F_0$.
\end{prop}
\begin{proof}
See Appendix \ref{app:top_affine}.
\end{proof}

\subsection{The finite-dimensional case: natural topologies}
Let us consider a finite-dimensional, real vector space $F_0$. By the Tychonoff theorem (\emph{e.g.}, see \cite{Schaefer}), there is a unique Hausdorff topology $\T_{0N}$ on $F_0$ making continuous the vector operations; this is called the \emph{natural topology} of $F_0$. Now let $F$ be an affine space over $F_0$. The translated topology of $\T_{0N}$, say $\T_N$, is called the \emph{natural topology} of $F$. If we use $\T_N$ and $\T_{0N}$, the sum $F\times F_0 \to F$, $(p,v)\mapsto p+v$ turns out to be continuous and, for each $p\in F$, the map $\theta_p\colon F\to F_0$, $q\mapsto q-p$ is a homeomorphism. The latter statement implies that $\T_N$ is first countable, \emph{i.e.}, there is a countable neighborhood base for each point $p\in F$ (this base can be chosen to be nested).

When we want to emphasize the dependence of the topology on $F_0$ and $F$, we write $\T_{0N}(F_0)$ and $\T_N(F)$. Furthermore, we write $F_0^N$ (resp., $F^N$) to indicate a vector space $F_0$ (resp., an affine space $F$) equipped with the natural topology. Finally, let $S\subset F$ be an affine subspace. In principle, we have on $S$: the induced topology $S\cap \T_N(F):=\{S\cap\Omega \mid \Omega\in\T_N(F)\}$; the natural topology $\T_N(S)$, defined regarding $S$ as an affine space in its own right. It turns out that $S\cap \T_N(F) = \T_N(S)$. Of course, $S^N$ will mean $S$ with this topology.

\subsection{Minkowski spacetime $M$ and its natural topology}\label{minkowski_def}
Let $M$ be the ($k$+1)-dimensional Minkowski spacetime; its associated vector space $M_0$ is a ($k$+1)-dimensional, real vector space endowed with a nondegenerate symmetric bilinear form $g$ of signature $(-,\underbrace{+,\dots,+}_{k\text{ times}})$.

As usual, a vector $v\in M_0$ is \emph{spacelike}, \emph{lightlike} or \emph{timelike} if $g(v,v)$ is, respectively, positive, null or negative. The \emph{light cone} with vertex $p\in M$ is:
\eq\label{lightcone}
C(p):=\{q\in M \mid g(q-p,q-p)=0\}\quad.
\eeq

An orthonormal frame of reference in $M$ is a pair $(o,(e_i)_{i=0,\dots,k})$ where $o\in M$ is called the origin and $(e_i)_{i=0,\dots,k}$ is an orthonormal base for $M_0$, that is a vector base for $M_0$ such that $g(e_i,e_j)=0$ for $i\neq j$, $g(e_0,e_0)=-1$, $g(e_i,e_i)=1$ for $i=1,\dots,k$.

Being finite dimensional, $M$ has its natural topology $\T_N$ (this is the ``standard topology'' mentioned in the Introduction). Similarly $M_0$ has the natural topology $\T_{0N}$.

\section{An introduction to the Zeeman topology on $M$}\label{sec:Zeeman_defined}

\subsection{Defining the topology}
\begin{defn}
In $M_0$, spacelike hyperplanes passing through the origin and straight timelike lines passing through the origin are called \emph{vector axes}. The set of vector axes is called $\mathcal{A}_0$. A set $p+A_0$, with $p\in M$ and $A_0\in\mathcal{A}_0$, is called an \emph{axis}. $\mathcal{A}$ will denote the set of axes.
\end{defn}

\begin{defn}\label{def:Zeeman}
The \emph{Zeeman topology} $\T_Z\equiv \T_Z(M)$ is the finest topology on $M$ which induces the affine space natural topology on every axis. $M$ with this topology will be written $M^Z$.
\end{defn}

\begin{lemma}\label{lemma:aperti}
Let $\Omega\subset M$. Then $\Omega$ is open in $M^Z$ $\Longleftrightarrow$ for every $A\in\mathcal{A}$, the set $\Omega\cap A$ is open in $A^N$.
\end{lemma}

\begin{proof}
The implication $\implies$ is trivial. To prove $\Longleftarrow$, let $\mathcal{C}$ be the collection of sets $\Omega\subset M$ such that $\Omega\cap A$ is open in $A^N$ for every $A\in\mathcal{A}$. We have the following:
\begin{itemize}
\item   $\emptyset$ and $M$ are in $\mathcal{C}$;
\item   given $A\in\mathcal{A}$ and a family $(\Omega_j)_{j\in J}$ where $\Omega_j\in\mathcal{C}$ for each $j$, we have $(\bigcup_j \Omega_j)\cap A = \bigcup_j\left(\Omega_j\cap A\right)$ and the latter is open in $A^N$ being a union of open sets;
\item   given a finite family $(\Omega_j)_{j\in J}$ where $\Omega_j\in\mathcal{C}$ for each $j$, we have $(\bigcap_j \Omega_j)\cap A = \bigcap_j(\Omega_j\cap A)$ and the latter is open in $A^N$ being a finite intersection of open sets.
\end{itemize}
So $\mathcal{C}$ is a topology; clearly, this implies $\mathcal{C}=\T_Z$.
\end{proof}

\begin{lemma}\label{lemma:chiusi}
Let $G\subset M$. Then $G$ is closed in $M^Z$ $\Longleftrightarrow$ for every $A\in\mathcal{A}$, the set $G\cap A$ is closed in $A^N$.
\end{lemma}
\begin{proof}
$G$ is closed in $M^Z$ if and only if $M\setminus G$ is open in $M^Z$, and by Lemma \ref{lemma:aperti} $M\setminus G$ is open in $M^Z$ if and only if for every axis $A\in\mathcal{A}$ the set $A\cap (M\setminus G)$ is open in $A^N$. Being $A\cap (M\setminus G)=A\setminus (G\cap A)$, the last condition is equivalent to closedness of $G\cap A$ in $A^N$.
\end{proof}

\subsection{Elementary facts on the Zeeman topology}
Since $\T_N \subset \T_Z$, it follows that $M^Z$ is Hausdorff. By the same reason, given a topological space $X$, the set $\mathcal{C}(M^Z,X)$ of continuous maps $f\colon M^Z\to X$ contains the set $\mathcal{C}(M^N,X)$ of continuous maps $f\colon M^N\to X$. Dually, the set $\mathcal{C}(X,M^N)$ contains the set $\mathcal{C}(X,M^Z)$.

In the sequel, the term \emph{light ray} means any lightlike straight line in $M$.

\begin{lemma}\label{lightray}
Any light ray $\lambda$ is a discrete subset of $M^Z$ (\emph{i.e.}, each point $p\in\lambda$ is isolated for $\lambda$).
\begin{proof}
For each $p\in\lambda$ and any open neighborhood $B\ni p$ in $M^N$, take the set $B':=\{B\setminus C(p)\}\cup \{p\}$. $B'$ is open in $M^Z$; in order to see this, let $A$ be an axis; if $A$ is passing through $p$ then $B'\cap A= B\cap A$, otherwise $B'\cap A=B\cap(M\setminus C(p))\cap A$; in either case $A$ meets $B'$ in an open set of $A^N$, so by Lemma \ref{lemma:aperti} $B'$ is open in $M^Z$. Since $B'\cap \lambda=\{p\}$, then the Zeeman topology $\T_Z$ induces on $\lambda$ the discrete topology.
\end{proof}
\end{lemma}

By a similar argument, one proves the following:

\begin{lemma}\label{isolated_vertex}
Consider the light cone $C(p)$ of any point $p$. Then, $p$ is an isolated point for $C(p)$ in $M^Z$.
\end{lemma}

\subsection{$\T_Z$ as a translated topology}

\begin{prop}\label{prop:tran_omeo}
$\T_Z$ is compatible with translations.
\end{prop}
\begin{proof}
Let $\Omega$ be open in $M^Z$ and $v\in M_0$. We claim that $T_v(\Omega)=\Omega+v$ is open in $M^Z$. In fact, for every $A \in \mathcal{A}$ we have $(\Omega+v)\cap A= (\Omega\cap (A-v))+v$ and $A-v\in\mathcal{A}$, so $\Omega\cap(A-v)$ is open in $(A-v)^N$ by hypothesis. But $T_v$ gives a homeomorphism in the natural topologies of the axes, so $(\Omega\cap(A-v))+v$ is open in $((A-v)+v)^N=A^N$. Since $T_v(\Omega)$ is open in $M^Z$ for each $v$ and $T_v^{-1}=T_{-v}$, it follows that every $T_v$ is a homeomorphism on $M^Z$.
\end{proof}
From Props. \ref{prop:top_affine} and \ref{prop:tran_omeo} it follows that $\T_Z$ is realizable as a translated topology. The natural candidate to give $\T_Z$ as a translated topology is $\T_{0Z}$, defined as follows:
\begin{defn}
$\T_{0Z}$ is the Zeeman topology (in the sense of Def. \ref{def:Zeeman}) of $M_0$, viewed as an affine space.\hfill $\diamond$
\end{defn}
In fact, we have:
\begin{prop}\label{prop:top0Z}
$\T_Z$ is the translated topology of $\T_{0Z}$.
\end{prop}
\begin{proof}
See Appendix \ref{app:def_Zeeman}.
\end{proof}
For the case $k=1$, we can exhibit on $M_0$ a different topology $\T_{0Z}'$ whose translated topology is $\T_Z$:
\begin{prop}\label{prop:top0Z'}
Let $k=1$, and denote by $\T_{0Z}'$ the finest topology on $M_0$ which induces the vector space natural topology on every vector axis. Then $\T_Z$ is the translated topology of $\T_{0Z}'$.
\end{prop}
\begin{proof}
See Appendix \ref{app:def_Zeeman}.
\end{proof}

\subsection{$\T_Z$ as a final topology}
Let $Y$ be a set; consider two families $(X_j)_{j\in J}$, $(f_j)_{j\in J}$ where, for each $j$, $X_j$ is a topological space and $f_j\colon X_j\to Y$ a map. The \emph{final topology} with respect to the spaces $(X_j)_{j\in J}$ and the maps $(f_j)_{j\in J}$ is the finest topology $\T^{\text{fin}}$ on $Y$ making all $f_j$ continuous. It is well known that:
\eq
\T^{\text{fin}}=\{E\subset Y \mid \forall j\in J \quad f_j^{-1}(E)\text{ is open in }X_j\}\quad.
\eeq
Given a topology on $Y$, the following are equivalent:
\begin{enumerate}
\item for every topological space $Z$ and every map $g\colon Y\to Z$:

$g$ is continuous $\Longleftrightarrow$ $\forall j\in J$,\; $g\circ f_j$ is continuous;
\item $Y$ has the final topology with respect to the spaces $(X_j)_{j\in J}$ and the maps $(f_j)_{j\in J}$.
\end{enumerate}
It is possible to describe $M^Z$ as a final topology. Let $\mathcal{A}$ be the set of axes and let us index it with itself. Now endow every axis $A$ of the family with its affine space natural topology, and consider the natural inclusion:

\begin{align*}
i_{A}\colon A&\hookrightarrow M\\
x &\mapsto x\quad.
\end{align*}
The following holds:

\begin{thm}\label{teor:M^Z_top_finale}
$\T_Z$ is the final topology with respect to the spaces $(A^N)_{A\in\mathcal{A}}$ and the maps $(i_A)_{A \in \mathcal{A}}$.
\end{thm}
\begin{proof}
By Lemma \ref{lemma:aperti}, $\Omega\subset M$ is open in $M^Z$ if and only if for every $A \in \mathcal{A}$ the set $\Omega\cap A$ is open in $A^N$. But $\Omega\cap A = i_{A}^{-1}(\Omega)$, so this means that $\Omega$ belongs to the final topology with respect to the spaces $(A^N)_{A\in\mathcal{A}}$ and the maps $(i_A)_{A\in \mathcal{A}}$.
\end{proof}
From item 1 above and Theorem \ref{teor:M^Z_top_finale} it follows the following corollary:

\begin{cor}\label{cor:restrizioni}
Given a topological space $T$, a map $f\colon M^Z\to T$ is continuous if and only if, for every $A\in\mathcal{A}$, the restriction $f\restriction_A$ is continuous on $A^N$.
\end{cor}

\subsection{Homeomorphism group of $M^Z$}
Let us recall the fact mentioned in the Introduction, giving the main motivation to study this topology.
\begin{thm}[Zeeman]
The homeomorphism group of $M^Z$ is generated by the Lorentz group, translations and dilatations.
\end{thm}
\begin{proof}
See \cite{Zeeman}.
\end{proof}

\subsection{A function continuous on $M^Z$, but not on $M^N$}

Let us fix a point $p$ and a (nonzero) timelike vector $e$. Let $h\colon [0,+\infty) \to \real$ be a continuous function such that $h(s) \to 1$ for $s \to +\infty$ and $h(0) = 0$. Given two exponents $\alpha, \beta > 0$, let us define $f\colon M\to \real$ setting:

\eq
f(x):=
\begin{cases}
h\left(\displaystyle{\frac{|g(e,x-p)|^{2\alpha + \beta}}{|g(x-p,x-p)|^\alpha}}\right) & x \notin C(p)\;, \\
1 & x \in C(p)\setminus\{p\}\;, \\
0 & x=p\;.
\end{cases}
\eeq
($C(p)$ is defined by Equation (\ref{lightcone}), Subsec. \ref{minkowski_def}). It is easy to verify that $f$ is continuous at each point of $M^N$ different from $p$, but not at $p$. However, we have the following:

\begin{prop}
$f$ is continuous on $M^Z$.
\end{prop}

\begin{proof}
For each axis $A$ not passing through $p$, $f\restriction_A$ is clearly continuous on $A^N$. Let us show the same happens for every axis $A$ through $p$. Of course, the nontrivial point is the continuity of $f\restriction_A$ at $p$. To prove this, let us represent the axis as $p+A_0$, with $A_0$ a vector axis. In both the timelike and the spacelike case, the function $v\in A_0 \mapsto |g(v,v)|^{1/2}$ is a norm on the vector space $A_0$. Clearly the function $v\mapsto g(e,v)$ is a linear form on $A_0$, so there is a constant $c>0$ such that $|g(e,v)|\leq c|g(v,v)|^{1/2}$ for all $v\in A_0$. This implies:
\eq
\frac{|g(e,x-p)|^{2\alpha + \beta}}{|g(x-p,x-p)|^\alpha} \leq c^{2\alpha + \beta}|g(x-p,x-p)|^{\beta/2}\to 0
\eeq
for $x\to p$ in $A^N$. This implies $f(x) \to h(0)=0=f(p)$ for $x \to p$. Then, by Corollary \ref{cor:restrizioni}, $f$ is continuous on $M^Z$.
\end{proof}

\section{Zeno sequences}\label{sec:Zeno}
Following Zeeman \cite{Zeeman}, we introduce the concept of Zeno sequence and give some example. This will be thoroughly used in the sequel, leading to a full characterization of compact sets in the 1+1 dimensional $M^Z$.
\begin{defn}
A \emph{Zeno sequence} is a sequence $(p_n)_{n \in \natr}$ of distinct points of $M$, converging to some point $p$ in $M^N$ but not in $M^Z$. The \emph{completed image of a Zeno sequence} is $\{p_n\}_{n\in\natr}\cup\{p\}$.
\end{defn}
\begin{exam}[Lightlike Zeno sequence]\label{esem:luce}
Consider a light ray $\lambda$. By Lemma \ref{lightray}, the Zeeman topology $\T_Z$ induces on $\lambda$ the discrete topology. Clearly every sequence of distinct points in $\lambda$ converging in $M^N$ to a point of $\lambda$ is a Zeno sequence. More generally, by Lemma \ref{isolated_vertex}, every sequence of distinct points in $C(p) \setminus \{p\}$ converging to $p$ in $M^N$ is a Zeno sequence.
\end{exam}
\begin{exam}[Axial Zeno sequence]\label{esem:assi}
Given $p\in M$, let us consider the set $\mathcal{A}_p$ of axes
passing through $p$ and choose a sequence $(p_n)_{n\in\natr}$ of
distinct points different from $p$, converging to $p$ in $M^N$ and
such that at most a finite number of them lie on a same axis
$A\in\mathcal{A}_p$. Then $(p_n)_{n\in\natr}$ is a Zeno sequence.
It suffices to show that\footnote{We use the symbol $|S|$ to
indicate the cardinality of a set $S$.} $|\{p_n\}_{n\in\natr}\cap
A|<\infty$ for every axis $A\in\mathcal{A}$ (so, not necessarily
passing through $p$); in fact, by Lemma \ref{lemma:chiusi} this
implies $\{p_n\}_{n\in\natr}$ is closed in $M^Z$, so $p$ is not a
limit point for $\{p_n\}_{n\in\natr}$. In order to get the thesis,
we suppose the set $\{n \mid p_n\in A\}$ to be an infinite
sequence $n_0< n_1 < n_2 \dots$ and derive a contradiction. In
fact, in this case $p=\lim_{j\to \infty} p_{n_j}$ in $M^N$; but
$A$ is closed in $M^N$, hence $p\in A$. So, $A\in \mathcal{A}_p$
and, for this reason, $A$ meets $\{p_n\}_{n\in\natr}$ finitely.
\end{exam}

\section{The 1+1 dimensional case. Subsets of $M$ on which $\T_N$ and $\T_Z$ induce the same
topology}\label{sec:1+1subsets}
In this section we assume $k=1$. We characterize the sets $\Sigma\subset M$ on which $\T_N$ and $\T_Z$ induce the same topology, \emph{i.e.}, such that $\Sigma \cap \T_Z = \Sigma\cap\T_N$.
\begin{prop}\label{prop:nozeno}
Let $\Sigma\subset M$, and consider the following statements:
\begin{enumerate}
\item[0.] $\Sigma$ contains no completed images of Zeno sequences.

\item[1.] $\forall p\in \Sigma$ there is an open neighborhood $W$ of $p$ in $M^N$ such that:
\eq W\cap \Sigma\cap C(p)=\{p\} \;.\eeq

\item[2.] $\forall p\in \Sigma$ there is an open neighborhood $U$ of $p$ in $M^N$ such that $U \cap \Sigma$ can be covered by finitely many axes passing through $p$\;.

\item[3.] $\forall p\in \Sigma$ and for every open neighborhood $S$ of $p$ in $M^Z$, there is an open neighborhood $T$ of $p$ in $M^N$ such that $T\cap \Sigma\subset S\cap \Sigma$\;.

\item[4.] $\Sigma\cap\T_Z = \Sigma\cap\T_N$\;.
\end{enumerate}
Then $\textit{0}\implies \textit{1,2,3}$ and $\textit{0}\Longleftrightarrow \textit{4}$.
\end{prop}
\begin{proof}\hfill
\begin{enumerate}
\item[$\textit{0}\Rightarrow \textit{1}$.] Suppose \textit{0}, and assume there is a point $p \in \Sigma$ not satisfying the claim \textit{1}, that is for every open neighborhood $W$ of $p$ in $M^N$ it is $W \cap \Sigma \cap C(p) \setminus \{p\}   \neq \emptyset$. Consider a nested neighborhood base $(B_n)_{n\in\natr}$ for $p$ in $M^N$ and inductively construct a sequence of integers $n_0< n_1 <\dots$ and a sequence of distinct points $p_k \in B_{n_k} \cap \Sigma \cap C(p) \setminus \{p\}$. Then $(p_k)_{k \in \natr}$ is a Zeno sequence as in Example \ref{esem:luce}, and $\{p_k\}_{k \in \natr} \cup \{p\} \subset \Sigma$, contradicting the hypothesis on $\Sigma$.

\item[$\textit{0}\Rightarrow \textit{2}$.] Suppose \textit{0}, and assume there is a point $p \in \Sigma$ not satisfying the claim \textit{2}; so, for every open neighborhood $U$ of $p$ in $M^N$ the set $U \cap \Sigma$ cannot be covered by finitely many axes passing through $p$. By the implication $\textit{0}\Rightarrow \textit{1}$, which we have just proven, there is an open neighborhood $W$ of $p$ in $M^N$ such that $W\cap \Sigma\cap C(p)=\{p\}$. Consider a nested neighborhood base $(B_n)_{n\in\natr}$ for $p$ in $M^N$; let $B_{n_0}\subset W$. Let $\mathcal{A}_p$ be the set of axes through $p$. By the initial assumption, for each $n\in\natr$ the set $\{ A\in\mathcal{A}_p \mid A\cap B_n\cap\Sigma\setminus \{p\}\neq\emptyset\}$ is infinite. Using this fact, construct a strictly increasing sequence $(n_j)_{j\in\natr}$ of integers, a sequence $(A_j)_{j\in\natr}$ of distinct axes and a sequence $(p_j)_{j\in\natr}$ of points such that $\forall j\in\natr$ $p_j\in A_j\cap B_{n_j}\cap\Sigma\setminus\{p\}$. From $p_j\in B_{n_j}$, it follows that $p_j\to p$ in $M^N$. If $A\in\mathcal{A}_p$ and $A\neq A_j$ for every $j\in\natr$, then $A\cap\{p_j\}_{j\in\natr}=\emptyset$; if $A=A_i$ for some $i$, then $A\cap\{p_j\}_{j\in\natr}=\{p_i\}$ (these statements depend essentially on the fact that the axes are 1-dimensional). In conclusion, each axis through $p$ meets $\{p_j\}_{j\in\natr}$ in one point at most; by the Example \ref{esem:assi}, this means $(p_j)_{j\in\natr}$ is a Zeno sequence and $\{p_n\}_{n \in \natr} \cup \{p\} \subset \Sigma$,    contradicting the hypothesis on $\Sigma$.

\item[$\textit{0}\Rightarrow \textit{3}$.]  Let $\Omega$ be an open neighborhood of $p\in \Sigma$ in $M^Z$. By the implication $\textit{0}\Rightarrow \textit{2}$, which we have just proven, there is an open neighborhood $U$ of $p$ in $M^N$ such that $U \cap \Sigma$ can be covered by finitely many axes $(A_j)_{j=1, \dots, J}$ passing through $p$. Thus $\Omega \cap U \cap \Sigma$ can also be covered by such a  family of axes. Consider a nested neighborhood base $(B_n)_{n\in\natr}$ for $p$ in $M^N$; then for every $j=1,\dots,J$ the family $(B_n\cap A_j)_{n\in\natr}$ is a nested neighborhood base for $p$ in $A_j^N$. Now, $\Omega\cap U$ is an open neighborhood of $p$ in $M^Z$, thus $\Omega \cap U \cap A_j$ is an open neighborhood of $p$ in $A_j^N$, so there is $B_{n_j}$ such that $B_{n_j}\cap A_j \subset \Omega \cap U \cap A_j$. Let $B:=B_m$ where $m:=\max\{n_1,\dots,n_J\}$, so that $B\cap A_j\subset \Omega\cap\ U \cap A_j$ for $j=1,\dots,J$. Since $\bigcup_{j=1}^J A_j$ covers $U\cap\Sigma$, it follows that $B\cap U\cap \Sigma \subset \Omega \cap U \cap  \Sigma\subset \Omega\cap \Sigma$ and the set $T:=B\cap U$, open in $M^N$, is    what we were looking for.

\item[$\textit{0}\Rightarrow \textit{4}$.] Let $\Omega$ be an open set of $M^Z$ meeting $\Sigma$. By the implication $\textit{0}\Rightarrow \textit{3}$, which we have just proven, for each $p\in \Omega\cap \Sigma$ there is an open neighborhood $T=T_p$ of $p$ in $M^N$ such that $T_p \cap \Sigma \subset \Omega \cap \Sigma$. Then $U:= \bigcup_{p \in \Omega \cap \Sigma} T_p$ is open in $M^N$ and such that $U \cap \Sigma = \Omega \cap \Sigma$. So every set which is open in $\Sigma \cap \T_Z$ is open in $\Sigma\cap\T_N$ too; the converse is trivial, for $\T_N \subset \T_Z$.

\item[$\textit{4}\Rightarrow \textit{0}$.] Assume $\{p_n\}_{n\in\natr} \cup \{p\} \subset \Sigma$, where $(p_n)_{n\in\natr}$ is a Zeno sequence converging to $p$. Then, there is an open neighborhood $\Omega$ of $p$ in $M^Z$, and a subsequence $(p_{n_k})_{k\in\natr}$ such that $\Omega \cap \{p_{n_k}\}_{k\in\natr} = \emptyset$. Since every open neighborhood $T$ of $p$ in $M^N$ contains points belonging to the sequence, it follows $T \cap \Sigma \neq \Omega \cap \Sigma$ and thus $\Sigma \cap \T_Z \neq \Sigma \cap \T_N$.
\end{enumerate}
\end{proof}
From the implication $\textit{0}\Rightarrow \textit{4}$ of Prop. \ref{prop:nozeno}, it follows the following corollary:
\begin{cor}\label{cor:fX->M cont}
Let $X$ be a topological space and let $f\colon X\to M$ be such that $f(X)$ contains no completed images of Zeno sequences; then $f\colon X \to M^Z$ is continuous if and only if $f\colon X\to M^N$ is continuous.
\end{cor}
\begin{proof}
If $f\colon X \to M^Z$ is continuous, then $f\colon X\to M^N$ is obviously continuous. Conversely, if $f\colon X \to M^N$ is continuous, let $\Omega \in \T_Z$. By the implication $\textit{0}\Rightarrow \textit{4}$ of Prop. \ref{prop:nozeno} there is $U\in\T_N$ such that $\Omega \cap f(X)=U \cap f(X)$. Now $f^{-1}(\Omega) = f^{-1}(\Omega\cap f(X)) = f^{-1}(U\cap f(X)) = f^{-1}(U)$, and $f^{-1}(U)$ is open in $X$ by hypotheses.
\end{proof}

\section{Compactness in $M^Z$}\label{sec:compact}
\subsection{Some general facts. A full characterization of compact subsets of $M^Z$ in the 1+1 dimensional case.}

First of all, from $\T_N \subset \T_Z$ it follows at once the following lemma:
\begin{lemma}\label{lemma:compatti_banale}
A compact subset of $M^Z$ is compact in $M^N$.
\end{lemma}

\begin{lemma}\label{lemma:succ}
Let $X$ be a Hausdorff topological space and let $(x_n)_{n\in\natr}$ be a sequence of distinct points of $X$ converging to $x$. Then $x$ is the unique limit point for the set $\{x_n\}_{n \in \natr}$. In particular, every $x_j$ is an isolated point for $\{x_n\}_{n\in\natr}$.
\end{lemma}
\begin{proof}
Elementary (for completeness, it is reported in Appendix \ref{app:succ}).
\end{proof}
\begin{lemma}\label{lemma:sottozen}
Every Zeno sequence admits a subsequence whose image is a non closed, discrete subset of $M^N$, closed in $M^Z$.
\end{lemma}
\begin{proof}
Let $(p_n)_{n\in\natr}$ be a Zeno sequence. By Lemma \ref{lemma:succ}, every $p_j$ is an isolated point of $\{p_n\}_{n\in\natr}$ in $M^N$, that is, $\T_N$ induces the discrete topology on the image of the sequence. The same is true for each of its subsets. In the sequel we show that there is a subsequence of $(p_n)_{n\in\natr}$ whose image is closed in $M^Z$, but not in $M^N$; this will conclude the proof.

First of all, consider any subsequence $(p_{n_k})_{k\in\natr}$. By Lemma \ref{lemma:succ} every $q\neq p$ is not a limit point for $\{p_{n_k}\}_{k\in\natr}$ in $M^Z$ ($\T_Z$ is finer than $\T_N$, hence every limit point with respect to $\T_Z$ is also a limit point with respect to $\T_N$). It follows that $\{p_n\}_{n\in\natr}$ is closed in $M^Z$ if and only if $p$ is not a limit point for $\{p_{n_k}\}_{k\in\natr}$ in $M^Z$. It remains to prove the existence of a subsequence $\{p_{n_k}\}_{k\in\natr}$ for which $p$ is not a limit point in $M^Z$. In fact, since $\{p_n\}_{n\in\natr}$ is not converging to $p$ in $M^Z$, there is an open neighborhood $U_p$ of $p$ in $M^Z$ such that for every $m\in\natr$ there is $n>m$ with $p_n\notin U_p$. Thus it is possible to inductively construct a subsequence $(p_{n_k})_{k \in \natr}$ such that $p_{n_k}\notin U_p$ for all $k$.
\end{proof}
\begin{thm}[Zeeman]\label{teor:zeeman}
A compact subset $K$ of $M^Z$ contains no images of Zeno sequences.
\end{thm}
\begin{proof}
Assume $K$ contains the image of a Zeno sequence. By Lemma \ref{lemma:sottozen} $K$ contains a nonclosed, discrete subset $Y$ of $M^N$, closed in $M^Z$. Being closed in a compact set, $Y$ is compact in $M^Z$. Since $Y$ is not closed in $M^N$, it must be $|Y|=\infty$. Since it is a discrete subset of $M^N$, for each $p\in Y$ there is an open neighborhood $U_p$ in $M^N$ such that $U_p\cap Y=\{p\}$. Therefore the family $(U_p)_{p\in Y}$ is a cover of $Y$ made of sets which are open in $M^N$, therefore open in $M^Z$, admitting no finite subcover. This contradicts the fact that $Y$ is compact in $M^Z$.
\end{proof}
Let us prove a stronger result for the 1+1 dimensional case.
\begin{thm}\label{teor:compatti1}
Let $k=1$; for a subset $K\subset M$, the following are equivalent:
\begin{enumerate}
\item $K$ is compact in $M^Z$.
\item $K$ is compact in $M^N$ and contains no completed images of Zeno sequences.
\end{enumerate}
\end{thm}
\begin{proof}\hfill
\begin{enumerate}
\item[$\textit{1}\Rightarrow \textit{2}$.] This follows from Lemma \ref{lemma:compatti_banale} and Theorem \ref{teor:zeeman}.
\item[$\textit{2}\Rightarrow \textit{1}$.] Let $K$ be a compact subset of $M^N$, containing no completed images of Zeno sequences. Consider a cover $(\Omega_\alpha)_{\alpha\in I}$ of $K$ by open subsets of $M^Z$. By the implication $\textit{0}\Rightarrow \textit{4}$ of Prop. \ref{prop:nozeno}, for each $\alpha$ there is an open set $\widehat{\Omega}_\alpha$ in $M^N$ such that $\widehat{\Omega}_\alpha\cap K = \Omega_\alpha\cap K$. Thus, $(\widehat{\Omega}_\alpha)_{\alpha\in I}$ is an open cover of $K$ in $M^N$. Since $K$ is compact in $M^N$, there is a finite subcover, say, $(\widehat{\Omega}_\alpha)_{\alpha\in J}$ with $J\subset I$; clearly, $(\Omega_\alpha)_{\alpha\in J}$ is a finite subcover extracted from $(\Omega_\alpha)_{\alpha\in I}$.
\end{enumerate}
\end{proof}

Sticking again to the 1+1 dimensional case, we characterize the compacts of $M^Z$ in terms of axes.
\begin{thm}\label{teor:compatti2}
If $k=1$ and $K\subset M$, the following are equivalent:
\begin{enumerate}
\item[1.] $K$ is compact in $M^Z$.
\item[3.] $K$ is covered by a finite family $(A_j)_{j=1,\dots,J}$ of axes such that for each $j=1,\dots,J$ the set $A_j\cap K$ is compact in $A_j^N$.
\end{enumerate}
\end{thm}
\begin{note}
In other words, every compact set of $M^Z$ is the union of finitely many compact sets of $M^N$, each lying on an axis.
\end{note}
\begin{proof}\hfill
\begin{enumerate}
\item[$\textit{3}\Rightarrow \textit{1}$.] If $K$ has a cover like in the claim \textit{3}, then, since $A_j \cap \T_N = A_j \cap \T_Z$, every $A_j\cap K$ is compact in $A_j\cap\T_Z$, hence in $M^Z$; finite unions of compact sets are compact.
\item[$\textit{1}\Rightarrow \textit{3}$.] Let $K\subset M^Z$ be compact. By Theorem \ref{teor:zeeman} and by the implication $\textit{0}\Rightarrow \textit{2}$ of Prop. \ref{prop:nozeno}, for each $p\in K$ there is an open neighborhood $U_p$ of $p$ in $M^N$ such that $U_p \cap K$ can be covered by a finite family of axes $(A_{p,j})_{j=1,\dots,J(p)}$ passing through $p$. The family $(U_p)_{p\in K}$ is a cover of $K$ made of open sets of $M^N$, from which a finite subcover $(U_{p_l})_{l=1,\dots L}$ can be extracted being $K$ compact in $M^N$; since $U_{p_l}\cap K$ is covered by the finite family $(A_{p_l,j})_{j=1,\dots,J(p_l)}$, it follows that $K$ is covered by the finite family $(A_{p_l,j})_{l=1,\dots,L\;;\; j=1,\dots,J(p_L)}$, as required by \textit{3}. Furthermore, since $K$ is compact in $M^N$ and every axis $A\in\mathcal{A}$ is closed in $M^N$, $A\cap K$ is a closed set contained in a compact set, hence it is compact in $M^N$. Being $A\cap K\subset A$, then $A\cap K$ is compact in $A^N$ as well. This holds, in particular, for all axes $A=A_{p_l,j}$, as required by \textit{3}.
\end{enumerate}
\end{proof}

\subsection{Weaker compactness properties}
We return to the case of arbitrary $k$.
\begin{lemma}\label{lemma:zenoaperti}
Every open set $\Omega$ of $M^Z$ contains completed images of Zeno sequences. More precisely, each point in $\Omega$ is the limit of a Zeno sequence whose image belongs to $\Omega$.
\end{lemma}
\begin{proof}
Given $p\in \Omega$, every axis $A$ passing through $p$ is such that $\Omega\cap A$ is an open neighborhood of $p$ in $A^N$. In particular, we use the timelike axes through $p$, which are 1-dimensional. Using this fact, it is easy to construct inductively a sequence of distinct points $p_n$ of $\Omega$ converging to $p$ in $M^N$, such that $p_n$ and $p_{n'}$ belong to distinct timelike axes for $n\neq n'$. For each axis $A$ through $p$, $\{p_n\}_{n\in\natr}\cap A$ consists of one point at most (in particular, is empty for $A$ spacelike). So, $(p_n)_{n\in\natr}$ is a Zeno sequence of the type in Example \ref{esem:assi}.
\end{proof}
\begin{defn}
A topological space $X$ is called \emph{locally compact} if each point has a compact neighborhood.
$X$ is called \emph{Lindel\"of} if every open cover has a countable subcover.
\end{defn}
\begin{prop}
$M^Z$ is neither locally compact nor Lindel\"of.
\end{prop}
\begin{proof}
Failure of local compactness: if there were an open subset $\Omega$ of $M^Z$ and a compact subset $K$ of $M^Z$ such that $\Omega\subset K$, then by Lemma \ref{lemma:zenoaperti} $K$ would contain images of completed Zeno sequences, contradicting Theorem \ref{teor:zeeman}.

Failure of the Lindel\"of property: any closed subset of a Lindel\"of topological space is Lindel\"of in the induced topology\footnote{If $G$ is closed in the Lindel\"of space $X$, any open cover of $G$ can be enlarged to an open cover of $X$ adding the open set $X\setminus G$. Extracting a countable subcover from this open cover of $X$ and discarding the set $X\setminus G$, one obtains a countable subcover of the open cover of $G$.}. Therefore, it suffices to find a closed subset of $M^Z$ which is non-Lindel\"of in the induced topology. An example is a light ray, since it is a closed discrete uncountable subset of $M^Z$.
\end{proof}

\section{Countability properties of $M^Z$}\label{sec:count}
Chosen an orthonormal frame of reference $(o,(e_i)_{i=0,\dots,k})$, every $p\in M$ is univocally identified by its coordinates $\{p^i\}_{i=0,\dots,k}$, such that $p=o+\sum_{i=0}^k p^i e_i$.

\subsection{Separability}
Clearly $M^N$ is separable (so are all finite-dimensional affine spaces endowed with their natural topology). A countable dense subset $Q$ of $M^N$ can be constructed by choosing an orthonormal frame of reference and defining $Q$ as the set of points in $M$ with rational coordinates.

For what concerns $M^Z$, we have the following proposition:

\begin{prop}\label{prop:separabile}
For every orthonormal frame of reference, the above-mentioned set $Q$ is also dense in $M^Z$. Thus $M^Z$ is separable.
\end{prop}
\begin{proof}
Let $(o,(e_i)_{i=0,\dots,k})$ be an orthonormal frame of reference and define $Q$ as above. The thesis follows if we show that every non empty open subset $\Omega$ of $M^Z$ meets $Q$. Let us consider any such $\Omega$; we claim there is $p\in\Omega$ with $p^0\in\rat$. To prove this, choose any $q\in\Omega$. If $q^0\in\rat$, we take $p=q$. If $q^0\in\real\setminus\rat$, we consider the straight timelike line $\{q + \lambda e_0 \mid \lambda \in \real \}$ passing through $q$. Since $\Omega$ is open in $M^Z$, there is an interval $(-\epsilon, \epsilon) \subset \real$ such that $\{q + \lambda e_0 \mid \lambda \in (-\epsilon, \epsilon)\} \subset \Omega$. Having picked a rational $r \in (q^0-\epsilon, q^0 + \epsilon)$, the point $p:=q+(r-q^0)e_0$ has coordinates $(r,q^1,\dots,q^k)$ and it belongs to $\Omega$.

Having found $p\in\Omega$ with $p^0\in\rat$, let us consider the spacelike hyperplane $\{p + \sum_{i=1}^k \lambda^i e_i \mid \lambda^i\in\real, \; i=1,\dots,k\}$. Since $\Omega$ is open in $M^Z$, there are $k$ intervals $(-\epsilon_i, \epsilon_i) \subset \real$ with $i=1,\dots,k$ such that $\{p + \sum_{i=1}^k \lambda^i e_i \mid \lambda^i\in (-\epsilon_i, \epsilon_i) , \; i=1,\dots,k \} \subset \Omega$. Having chosen $k$ rationals $r^i\in (p^i-\epsilon_i, p^i + \epsilon_i)$, the point $s:=p+(r^i - p^i)e_i$ has rational coordinates $(p^0,r^1,\dots,r^k)$, so it belongs to $Q\cap\Omega$.
\end{proof}

\begin{cor}\label{cor:card_continue}
The cardinality of the set $\mathcal{C}(M^Z,\real)$ of all real continuous functions on $M^Z$ is at most equal to $2^{\aleph_0}$.
\end{cor}
\begin{proof}
This is a general property of separable spaces: in general, given a topological space $X$, any continuous function $f\colon X\to \real$ is determined by its values on any dense subset $D$ of $X$ (\emph{e.g.}, see \cite{Dugundji}); when $X$ is separable, we can choose $D$ to be countable. Therefore the cardinality of the set of real continuous functions on $X$ is at most equal to $|\mathbb{R}|^{|D|} = \left(2^{|\natr|}\right)^{|\natr|} = 2^{|\natr|^2} = 2^{|\natr|} = 2^{\aleph_0}$.
\end{proof}

\subsection{First countability}
\begin{prop} $M^Z$ is not first countable at any point.\end{prop}
\begin{note}
This was claimed, but not proven in \cite{Zeeman}.
\end{note}
\begin{proof}
Assume the claim is false: for some $p\in M$ there is a countable neighborhood base $(U_n)_{n \in \natr}$ in $M^Z$. We can take the $U_n$ to be open and nested. We can inductively construct a sequence of distinct points $(p_j)_{j\in\natr}$ different from $p$, a sequence of distinct timelike axes $(A_j)_{j\in\natr}$ through $p$ and a sequence $n_0< n_1 <n_2\dots$ such that $p\neq p_j\in A_j\cap U_{n_j}$ for each $j$. The same argument in the proof of Lemma \ref{lemma:zenoaperti} shows that $(p_j)_{j\in\natr}$ is an axial Zeno sequence; this implies that $\{p_j\}_{j\in\natr}$ is closed in $M^Z$ (see Example \ref{esem:assi}). Then $U:=U_0\setminus \{p_j\}_{j\in\natr}$ is an open neighborhood of $p$ in $M^Z$. For each $n$ we have $U_n\not\subset U$: in fact, taken $j$ such that $n_j> n$, we have $U\not\ni p_j$ and $p_j\in U_{n_j}\subset U_n$. The result $U_n\not\subset U$ for each $n$ contradicts the initial assumption that $(U_n)_{n \in \natr}$ is a neighborhood base for $p$.
\end{proof}

\section{$M^Z$ is not normal}\label{sec:normality}
\begin{defn}
A topological space $X$ is called \emph{normal} if for each pair of disjoint closed sets $F,G\subset X$ there is a pair of disjoint open sets $U,V\subset X$ such that $F\subset U$, $G\subset V$.
\end{defn}

\begin{defn}
Given a topological space $X$ and a pair of disjoint closed sets $F,G\subset X$, a \emph{Urysohn function for $F$ and $G$} is a continuous function $f\colon X\to\real$ such that $f(F)=\{0\}$ and $f(G)=\{1\}$.\hfill $\diamond$
\end{defn}
A well-known lemma holds:
\begin{Uri}
A topological space $X$ is normal if and only if for each pair of disjoint closed sets $F,G\subset X$ there is a Urysohn function for $F$ and $G$.
\end{Uri}
\begin{proof}
\emph{E.g.}, see \cite{Dugundji}.
\end{proof}
\begin{thm}
$M^Z$ is not normal.
\end{thm}
\begin{note}
This result, with a sketch of the proof, can be found in \cite{Zeeman}. Here we give a fully explicit proof, based on the Urysohn Lemma.
\end{note}
\begin{proof}
Let us fix a light ray $\lambda$. Since $\lambda$ is a closed discrete subset of $M^Z$, it follows that each subset $G \subset \lambda$ is closed in $M^Z$. If $M^Z$ were normal, for each subset $G\subset \lambda$ there would be a Urysohn function for the closed, disjoint subsets $G$ and $\lambda\setminus G$, that is a continuous function $f\colon M^Z\to\real$ such that $f(G)=\{0\}$ and $f(\lambda \setminus G) =\{1\}$. Thus there would be at least as many real continuous functions on $M^Z$ as the subsets of $\lambda$. But $|\lambda|=2^{\aleph_0}$, so the cardinality of the set $\mathcal{C}(M^Z,\real)$ would be at least $2^{2^{\aleph_0}}$. This contradicts Corollary \ref{cor:card_continue}, hence $M^Z$ is not normal.
\end{proof}

\section{The 1+1 dimensional case. Path con\-nect\-ed\-ness and loop homotopy in
$M^Z$} \label{sec:homotopy}
In this section we assume $k=1$.

\subsection{Path connectedness}
\begin{prop}\label{prop:archiconnesso}
$M^Z$ is path connected.
\end{prop}
\begin{proof}
Given a pair of points $p,q\in M$, we must show that there is a continuous map $\gamma\colon [0,1] \to M^Z$ with $\gamma(0)=p$ and $\gamma(1)=q$; hereafter, any such map will be referred to as a $Z$-path from $p$ to $q$. If $q\notin C(p)$ then it suffices to choose any continuous map $\gamma\colon [0,1]\to A^N$ from $p$ to $q$, where $A$ is an axis passing through $p$ and $q$: from Corollary \ref{cor:fX->M cont} it follows that $\gamma$ is also a $Z$-path from $p$ to $q$. If $q\in C(p)$, pick a $r\in M$ different from $p$ and $q$ belonging neither to $C(p)$ nor to $C(q)$; there is a $Z$-path $\gamma_1$ from $p$ to $r$, and there is a $Z$-path $\gamma_2$ from $r$ to $q$; composing $\gamma_1$ with $\gamma_2$ and adjusting the ``traveling time'' one gets a $Z$-path from $p$ to $q$.
\end{proof}
\begin{cor}
$M^Z$ is connected.
\end{cor}

\subsection{Nontriviality of $\pi_1(M^Z)$}
For each path-connected topological space $X$, we denote by $\pi_1(X)$ the first homotopy group of $X$.
\begin{thm}
$\pi_1(M^Z)$ is nontrivial and possesses uncountably many subgroups isomorphic to $\integ$. In particular, $M^Z$ is not simply connected.
\end{thm}

\begin{proof}
We fix a base point $o\in M$, once and for all, and consider $Z$-loops based at $o$, \emph{i.e.}, continuous maps $\gamma\colon [0,1] \to M^Z$ with $\gamma(0)=\gamma(1)=o$. Homotopies in $M^Z$ for such loops will be referred to as $Z$-homotopies. $N$-loops and $N$-homotopies are defined similarly, in terms of $M^N$.

Let us take two ordered pairs of vectors $(t_1,s_1)\neq(t_2,s_2)$, where $t_1$, $t_2$ are timelike and $s_1$, $s_2$ are spacelike. Consider the following two maps, $i=1,2$:
\eq
\gamma_i\colon [0,1]\to M\qquad
\gamma_i(u)=
\begin{cases}
o + 4u t_i & 0\leq u \leq \frac{1}{4} \\
o + t_i + (4u - 1) s_i & \frac{1}{4}\leq u \leq \frac{1}{2} \\
o + (3 - 4u) t_i + s_i & \frac{1}{2}\leq u \leq \frac{3}{4} \\
o + 4(1 - u) s_i & \frac{3}{4}\leq u \leq 1 \\
\end{cases}
\eeq
(essentially, $\gamma_i$ is made of four straight line segments, tracing the sides of a parallelogram). Note that both $\gamma_i$, $i=1,2$, are $Z$-loops by Corollary \ref{cor:fX->M cont}, since each $\gamma_i$ is continuous in $M^N$ and $\gamma_i([0,1])$ is covered by finitely many (in fact, four) axes. We claim there is no $Z$-homotopy of loops (with base point $o$) taking $\gamma_1$ to $\gamma_2$. For let $H\colon [0,1]^2 \to M^Z$ be such a homotopy: then $H$ would also be an $N$-homotopy of loops (with base point $o$), since $\T_Z$ is finer than $\T_N$. Let us call $R_1$, $R_2$ the compact parallelograms in $M^N$ whose boundaries are $\gamma_1([0,1])$ and $\gamma_2([0,1])$ respectively, \emph{i.e.}, $R_i=\{p\in M\mid p=o+\lambda t_i + \mu s_i, \;\lambda,\mu\in [0,1]\}$, $i=1,2$. Since $(t_1,s_1)\neq(t_2,s_2)$, we have $R_1\neq R_2$, so at least one of the sets $\mathring{R_1}\setminus R_2$, $\mathring{R_2}\setminus R_1$ is not empty ($\mathring{S}$ means the interior of the set $S$ in the natural topology). Let $\mathring{R_1}\setminus R_2 \neq \emptyset$ and take a point $p\in\mathring{R_1}\setminus R_2$. Then $p\in H([0,1]^2)$, otherwise $H$ would be an $N$-homotopy between $\gamma_1$ and $\gamma_2$ in the punctured plane $M^N\setminus\{p\}$, and this is absurd since $\gamma_1$ winds around $p$ while $\gamma_2$ does not. So $\mathring{R_1}\setminus R_2 \subset H([0,1]^2)$. Now, $[0,1]^2$ is compact, so $H([0,1]^2)$ is compact in $M^Z$, but this is absurd since $\mathring{R_1}\setminus R_2$ is open in $M^N$, hence in $M^Z$, and by Lemma \ref{lemma:zenoaperti} it contains the image of a Zeno sequence, contradicting Theorem \ref{teor:zeeman}.

Calling $\mathcal{R}$ the set of all ordered pairs $(t,s)$ where $t$ is a timelike vector and $s$ is a spacelike vector, we associate to each pair $(t,s)$ a $Z$-loop $\gamma_{ts}$ with base point $o$ as in the above argument. Writing $[\gamma_{ts}]$ for the $Z$-homotopy class of $\gamma_{ts}$, we have then proven that the map $\mathcal{R}\to\pi_1(M^Z)$, $(t,s)\mapsto[\gamma_{ts}]$ is injective.

Now, consider $\gamma_{ts}\cdot\gamma_{ts}\equiv \gamma_{ts}^2$, where the dot means composition of loops in the homotopy group sense. Then $\gamma_{ts}^2$ is not $Z$-homotopic to $\gamma_{ts}$. For let $H$ be a $Z$-homotopy between $\gamma_{ts}^2$ and $\gamma_{ts}$. Then $H$ would also be an $N$-homotopy. Take the compact parallelogram $R_{ts}$ in $M^N$ whose boundary is $\gamma_{ts}([0,1])$: each $p\in \mathring{R}_{ts}$ is also in $H([0,1]^2)$, otherwise $H$ would be an $N$-homotopy between $\gamma_{ts}^2$ and $\gamma_{ts}$ in the punctured plane $M^N\setminus\{p\}$, which is absurd. In conclusion $\mathring{R}_{ts}\subset H([0,1]^2)$; again this is absurd since $\mathring{R}_{ts}$, being open in $M^Z$, contains the image of a Zeno sequence, while $H([0,1]^2)$ is compact in $M^Z$. Analogously $\gamma_{ts}^n$ is not $Z$-homotopic to $\gamma_{ts}^m$ for any integer $n\neq m$.

This shows that the map $\varphi_{ts}\colon \{[\gamma_{ts}^n]\mid n\in\integ\}\to\integ$, $[\gamma_{ts}^n]\mapsto n$ is a bijection. Moreover, since $[\gamma_{ts}^n][\gamma_{ts}^m]=[\gamma_{ts}^n\cdot\gamma_{ts}^m]=[\gamma_{ts}^{n+m}]$, it follows that $\varphi_{ts}$ is a group isomorphism and the claim is proven.
\end{proof}

\section*{ACKNOWLEDGMENTS}
%\addcontentsline{toc}{section}{Acknowledgment}
I would like to thank Professor Livio Pizzocchero for providing encouragement, suggestions and invaluable guidance, always with the utmost patience, generosity and goodwill. I would also like to thank Professor Gregory L. Naber for very kind comments and advices.

\begin{appendices}
\section{Proof of Prop. \ref{prop:top_affine}}\label{app:top_affine}
In this appendix we often use this obvious statement: a bijective map $\psi$ between two topological spaces is a homeomorphism if and only if $\psi$ and $\psi^{-1}$ are open (\emph{i.e.}, they map open sets into open sets).
Before proving Prop. \ref{prop:top_affine} we need the following Lemma:
\begin{lemma}\label{lemma:tran_top}
Let $\T$ be the translated topology on $F$ of a topology $\T_0$ on $F_0$. If for every $v\in F_0$ the vector translation $T_{0v}\colon F_0\to F_0$, $w\mapsto w+v$ is a homeomorphism, then for any $p\in F$ the bijection $\theta_p\colon F\to F_0$, $q\mapsto q-p$ is a homeomorphism.
\end{lemma}
\begin{note}
It is easy to see that if the vector sum $F_0\times F_0 \to F_0$, $(v,w)\mapsto v+w$ is continuous, then every vector translation $T_{0v}$ is a homeomorphism.
\end{note}
\begin{proof}[Proof of Lemma]
First we prove the map $\theta_p$ is open. Let $\Omega\in\T$. Then, by definition of translated topology, we can write $\Omega=\bigcup_{q\in\Omega} (q+O_q)$ where $O_q\in\T_0$, $0\in O_q$. Then $\theta_p(\Omega) = \Omega-p = \bigcup_{q\in\Omega}(O_q+q-p)$. Now, since every vector translation $T_{0v}\colon F_0\to F_0$, $w\mapsto w+v$ is a homeomorphism, it follows that $O_q+q-p \in \T_0$ (because $q-p\in F_0$). Then $\theta_p(\Omega)\in \T_0$ being a union of open sets.

Now we prove that $\theta_p^{-1}\colon F_0\to F$, $v\mapsto p+v$ is open. Let $O\in\T_0$. Then for every $v\in O$ there is an open neighborhood of $v$, say $J_v$, such that $J_v\subset O$. Since $T_{0v}$ is a homeomorphism, it follows that $J_v-v$ is open; moreover, $0\in J_v-v$. Now for every $q\in \theta_p^{-1}(O)=p+O$ we put $O_q:=J_v-v$ where $v=q-p$. By construction, $O_q$ is an open neighborhood of zero in $F_0$. Since $q+O_q=p+v+J_v-v=p+J_v\subset p+O$, it follows that $\theta_p^{-1}(O) \in\T$ by definition of translated topology.
\end{proof}

\begin{proof}[Proof of Prop. \ref{prop:top_affine}]\hfill
\begin{itemize}
\item[$\textit{2}\Rightarrow\textit{1}$.] Let $\T$ be the translated topology of $\T_0$. Let us consider any translation $T_v$ of $F$, and prove it is open. In fact, let $\Omega \in \T$ and consider $T_v(\Omega)=\Omega+v$. For every $p\in \Omega$ $\exists O_p\in\T_0$ such that $0\in O_p$ and $p+O_p\subset\Omega$; then, for every $q\in\Omega+v$, $q+O_{q-v}=q-v+v+O_{q-v}\subset\Omega+v$. This shows $T_v(\Omega) \in \T$. Since $T_v^{-1}=T_{-v}$, $T_v^{-1}$ is open as well.

\item[$\textit{1}\Rightarrow\textit{2}$.] Suppose $T_v$ is a homeomorphism for every $v\in F_0$. Choose a point $p\in F$ and consider the topology $\T_{0p}$ on $F_0$ obtained transporting $\T$ via the bijection $F\to F_0$, $q\mapsto q-p$. In fact $\T_{0p}$ does not depend on $p$: in order to see it, let us choose another point $p'$ to construct the bijection $F\to F_0, q\mapsto q-p'$. Then:
\eq
O\in\T_{0p} \Leftrightarrow p+O\in\T \Leftrightarrow p+O+p'-p\in\T \Leftrightarrow p'+O\in\T \Leftrightarrow O\in\T_{0p'}
\eeq
where the second equivalence holds since $T_{p'-p}$ is a homeomorphism. So $\T_{0p}$ is independent of $p$ and can be called $\T_0$. Now let $\T'$ be the translated topology of $\T_0$; in the following the equality $\T'=\T$ is proven. Take $\Omega\in\T$. Then $\Omega-p\in\T_0$ for every $p\in F$ by definition of $\T_0$. In particular, for every $p\in \Omega$ the set $O_p:=\Omega-p\in\T_0$ is a neighborhood of $0\in F_0$ such that $p+O_p=p+\Omega-p=\Omega$, so $\Omega\in\T'$. Summing up, $\T\subset \T'$. For the inverse inclusion: take $\Omega\in\T'$. Then $\Omega=\bigcup_{p\in \Omega}(p+O_p)$ with a suitable $O_p\in\T_0$. This implies $p+O_p\in\T$ by definition of $\T_0$. Then $\Omega$ is a union of open sets of $\T$, so $\Omega\in\T$. In conclusion, $\T'=\T$, \emph{i.e.}, $\T$ is the translated topology of $\T_0$.
\end{itemize}
\emph{Uniqueness part of the claim, under the hypothesis 1}. In order to prove the uniqueness part of the claim, let $\T_0$ and $\T_0'$ be two topologies on $F_0$ whose translated topology is $\T$ and such that every vector translation is a homeomorphism for both. Using Lemma \ref{lemma:tran_top} one easily proves that $O\in\T_0 \Longleftrightarrow p+O \in \T$ and $O\in\T_0' \Longleftrightarrow p+O \in \T$; so $O\in\T_0 \Longleftrightarrow O \in \T_0'$ and the full claim is proven.
\end{proof}

\section{Proofs of Props. \ref{prop:top0Z} and \ref{prop:top0Z'}}\label{app:def_Zeeman}
\begin{proof}[Proof of Prop. \ref{prop:top0Z}]
By Prop. \ref{prop:tran_omeo}, every vector translation $T_{0v}\colon M_0\times M_0\to M_0$, $(w,v)\mapsto w+v$ is a homeomorphism with respect to the topology $\T_{0Z}$. Now proceed as in the proof of Prop. \ref{prop:top_affine}, part $\textit{1}\Rightarrow \textit{2}$.
\end{proof}
\begin{proof}[Proof of Prop. \ref{prop:top0Z'}]
As usual, let $\T_N$ denote the natural topology of $M$. We recall that $k=1$. By definition,
\eq
\T_Z= \left\{ \Omega \subset M \mid \forall A\in\mathcal{A} \quad \exists U\in\T_N  \quad \text{s.t.} \quad \Omega \cap A = U \cap A\right\}\;.
\eeq
On the other hand, the translated topology of $\T_{0Z}'$ is:
\eq
\T_Z':= \left\{ \Omega \subset M \mid \forall p\in\Omega \quad \exists O\in\T_{0Z}' \quad \text{s.t.} \quad O\ni 0,\; p+O\subset\Omega\right\}\;.
\eeq
We claim that $\T_Z=\T_Z'$. First we take $\Omega\in\T_Z$, and check that $\Omega\in \T_Z'$. Given $p\in\Omega$, let us consider the set $\mathcal{A}_0$ of vector axes. For each $A_0\in\mathcal{A}_0$ there is $U_{A_0}\in\T_N$ such that $\Omega \cap(p+A_0)=U_{A_0}\cap(p+A_0)$; this implies $U_{A_0}-p\ni 0$. Now let:
\eq\label{def_O}
O:=\bigcup_{A_0\in\mathcal{A}_0}(U_{A_0}-p)\cap A_0 \quad ;
\eeq
For every $A_0'\in\mathcal{A}_0$ the following holds:
\eq
\begin{aligned}
O\cap A_0' &= \left(\bigcup_{A_0\in\mathcal{A}_0}(U_{A_0}-p)\cap A_0\right)\cap A_0'\\
&= \bigcup_{A_0\in\mathcal{A}_0}\left[(U_{A_0}-p)\cap A_0\cap A_0'\right]\\
&=(U_{A_0'}-p)\cap A_0'\quad.
\end{aligned}
\eeq
The last equality follows from the fact that, in dimension 1+1, $A_0\cap A_0'=\{0\}$ for $A_0'\neq A_0$. But $(U_{A_0'}-p)\cap A_0'$ is open in $A_0'^N$ since $U_{A_0'}-p$ is homeomorphic to the open subset $U_{A_0'}\in \T_N$. Therefore $O\in\T_{0Z}'$. From Equation (\ref{def_O}), it is also clear that $O\ni 0$. Finally, we note that:
\eq
\begin{aligned}
p+O &= p+\bigcup_{A_0\in\mathcal{A}_0}(U_{A_0}-p)\cap A_0\\
&=\bigcup_{A_0\in\mathcal{A}_0}\left\{p+\left[(U_{A_0}-p)\cap A_0\right]\right\}\\
&=\bigcup_{A_0\in\mathcal{A}_0}\left(p+U_{A_0}-p\right)\cap \left(p+A_0\right)\\
&=\bigcup_{A_0\in\mathcal{A}_0}U_{A_0}\cap \left(p+A_0\right)\\
&=\bigcup_{A_0\in\mathcal{A}_0}\Omega\cap \left(p+A_0\right)\subset\Omega\quad.
\end{aligned}
\eeq
This lets us conclude that $\Omega\in \T_Z'$.

Now we take $\Omega\in\T_Z'$, and check that $\Omega\in\T_Z$. Let $A\in\mathcal{A}$; we must prove that $\Omega\cap A=U\cap A$ for some $U\in\T_N$; this is trivial if $\Omega\cap A=\emptyset$, so in the sequel we assume $\Omega\cap A\neq\emptyset$. Let us denote by $A_0$ the unique vector axis such that $A=p+A_0$ for any $p\in A$. Now, consider any $p\in\Omega\cap A$; then by the assumption $\Omega\in\T_Z'$, there is $O_p\in\T_{0Z}'$ such that $O_p\ni 0$, $p+O_p\subset \Omega$. Furthermore, $O_p\in\T_{0Z}'$ implies $O_p\cap A_0=U_p\cap A_0$ for some $U_p\in\T_{0N}$. Let:
\eq
U:=\bigcup_{p\in\Omega\cap A}(p+U_p)\quad.
\eeq
We have $p+U_p\in\T_N$ for all $p$, so $U\in\T_N$. Furthermore:
\eq
\begin{aligned}
U\cap A&=\bigcup_{p\in\Omega\cap A}(p+U_p)\cap A\\
&=\bigcup_{p\in\Omega\cap A}(p+U_p)\cap(p+A_0)\\
&=\bigcup_{p\in\Omega\cap A}p+(U_p\cap A_0)\\
&=\bigcup_{p\in\Omega\cap A}p+(O_p\cap A_0)\\
&=\bigcup_{p\in\Omega\cap A}(p+O_p)\cap A\\
&=\left(\bigcup_{p\in\Omega\cap A}p+O_p\right)\cap A=\Omega\cap A\quad.
\end{aligned}
\eeq
\end{proof}

\section{Proof of Lemma \ref{lemma:succ}}\label{app:succ}

\begin{proof}
Let $y\in X$, $y\neq x$. Since $X$ is Hausdorff, there are two disjoint open neighborhoods $U_x$ and $U_y$ of $x$ and $y$ respectively. As the sequence converges to $x$, $U_x$ contains all but finitely many points of the sequence. Let $J\subset\natr$ be the finite set indexing all points of the sequence not belonging to $U_x$. If $y\notin\{x_n\}_{n\in\natr}$, the set $(X\setminus \{x_j\}_{j\in J}) \cap U_y$ is an open neighborhood of $y$ not containing points of the sequence, thus $y$ is not a limit point for $\{x_n\}_{n\in\natr}$. If $y=x_k$ for some $k\in\natr$, the set $(X\setminus \{x_j\}_{j\in J,\;j\neq k}) \cap U_y$ is an open neighborhood of $x_k$ not containing points of the sequence different from $x_k$. Thus $x_k$ is an isolated point for $\{x_n\}_{n\in\natr}$.
\end{proof}
\end{appendices}

\end{document}